\documentclass[doublecol,figures]{epl2}

\usepackage{amssymb,amsmath}
\usepackage{epsf,colordvi,graphicx,color,amsbsy}
\usepackage[T1]{fontenc}
\usepackage{ulem}

\title{Experimental study of the inverse cascade in gravity wave turbulence}
\author{L. Deike\thanks{Corresponding author: \email{luc.deike@univ-paris-diderot.fr}}, C. Laroche \and E. Falcon}
\shortauthor{L. Deike, C. Laroche \and E. Falcon}
\institute{Univ Paris Diderot, Sorbonne Paris Cit\'e, MSC, UMR 7057 CNRS, F-75 013 Paris, France, EU}

\abstract{
We perform experiments to study the inverse cascade regime of gravity wave turbulence on the surface of a fluid. Surface waves are forced at an intermediate scale corresponding to the gravity-capillary wavelength. In response to this forcing, waves at larger scales are observed. The spectrum of their amplitudes exhibits a frequency-power law at high enough forcing. Both observations are ascribed to the upscale wave action transfers of gravity wave turbulence. The spectrum exponent is close to the value predicted by the weak turbulence theory. The spectrum amplitude is found to scale linearly with the mean injected power. We measure also the distributions of the injected power fluctuations in the presence of upscale (inverse) transfers or in the presence of a downscale (direct) cascade in gravity wave turbulence.}
 

\begin{document}
\maketitle

\section{Introduction}
Waves on ocean surface is the most common example of wave turbulence. Wave turbulence concerns the dynamical and statistical study of a field of dispersive waves in nonlinear interaction. It is a rather universal phenomenon since it occurs in various physical contexts including geophysics, plasma physics, non linear optics or solid state physics \cite{ZakharovBook}. One of the most important result of wave turbulence theory is the existence of out-of-equilibrium stationary solutions for the wave spectrum that follow Kolmogorov-like cascades of flux of conserved quantities \cite{ZakharovBook}.  This cascade type behavior is similar to two-dimensional turbulence ones where both a direct cascade of enstrophy (rms vorticity) and an inverse cascade of energy occur \cite{KraichnanBatchelor,Rotgers98}. In wave turbulence, the cascades are governed by the nonlinear interaction process between waves: for a 4-wave process (as for gravity surface waves) the energy and the wave action are conserved \cite{ZakharovBook,McGoldrick}, whereas for a 3-wave process (as for capillary waves) only the energy is conserved \cite{ZakharovBook}. Frequency-power law solutions for gravity wave spectrum then exist corresponding to either a constant flux of energy from large to small scale (direct cascade) or a constant flux of wave action from small to large scale (inverse cascade) \cite{Zakharov82}. The direct gravity cascade has been observed in open seas \cite{Toba}, in well-controlled laboratory experiment \cite{Falcon07a, Nazarenko10}, and in numerical simulations \cite{Dyachenko04}. Existence of an inverse cascade in gravity wave turbulence has been confirmed recently using numerical simulations \cite{Annenkov06,Korotkevitch08}. These simulations notably show the formation of both inverse and direct gravity cascades without interacting each other and of the ``condensation'' of waves at large scale as an analogous of Bose-Einstein condensation in condensed matter physics \cite{Korotkevitch08}. Experiments related to the observation of an inverse cascade in wave turbulence are scarce. It concerns either non linear optics \cite{Bortolozzo09} or heat wave turbulence in a superfluid \cite{Ganshin08}. To our knowledge, no observation of an inverse cascade in gravity wave turbulence has been reported so far.

In this letter, we report the observation of upscale transfers in gravity wave turbulence. The energy injected into the waves at an intermediate scale (corresponding to the gravity-capillary length) generates waves of larger and larger scales due to the nonlinear wave interactions. A stationary state is reached and presents properties close to the inverse cascade of wave action predicted by wave turbulence theory \cite{ZakharovBook}. We show that the large scale cut-off of this upscale transfer regime is related to the horizontal finite size of the set-up. We characterized this regime by measuring the scaling of the wave amplitude spectrum with the frequency scale and with the power injected into the waves.  The probability density of the injected power fluctuations is also measured in the presence or in the absence of the upscale transfer regime of gravity wave turbulence.

\section{Experimental set-up}
The experimental setup is similar to the one used in \cite{Falcon07a}. It consists in a square plastic vessel, $L=20$ cm side, filled with mercury up to a height $h=18 $ mm. The properties of mercury are: density, $\rho = 13.5 \times10^{3}$ kg/m$^3$, kinematic viscosity, $\nu= 10^{-7}$ m$^2$/s and surface tension $\gamma=0.4$ N/m. Mercury is used because of its low kinematic viscosity. Surface waves are generated by a rectangular plunging wave maker (13 cm in length and 3.5 cm in height)  driven by an electromagnetic vibration exciter. The crossover frequency between gravity and capillary linear waves is $f_{gc}=\frac{1}{2\pi}\sqrt{2g/l_c}\simeq 17$ Hz with $g=9.81$ m/s$^2$ the acceleration of the gravity, and $l_c=\sqrt{\gamma/(\rho g)}$ is the capillary length \cite{Falcon07a}. Gravity waves thus occurs for frequency $ f < f_{gc}$ whereas capillary waves occurs for $f > f_{gc}$. The wave maker is driven around $f_{gc}$ in order to generate small scale gravity waves to be able to observe an upscale transfer regime from this small scale to larger ones. The wave maker is either driven sinusoidally at a frequency $f_p = 19$ Hz close to $f_{gc}$ or with a random noise (in amplitude and frequency) band-pass filtered around $f_p \pm 3$ Hz unless otherwise stated. The depth $H$ of the wave maker immersion is varied in a range $9 \leq H \leq 17$ mm. The amplitude of the surface waves $\eta(t)$ at a given location is measured by a capacitive wire gauge plunging perpendicularly to the fluid at rest \cite{Falcon07a}. The frequency cut-off of this probe is near 400 Hz. The signal $\eta(t)$ is recorded during 500 s using an acquisition card with a 2 kHz sampling rate. The instantaneous injected power into the fluid $I(t)$ is given by the product of the wave maker velocity $V(t)$ and the force $F(t)$ applied by the vibration exciter to the wave maker \cite{Falcon07a}. The mean injected power is thus $\langle I \rangle \equiv \langle F(t)V(t) \rangle$ where $\langle \cdot \rangle$ denotes a time average. $\sigma_F$ and $\sigma_V$ will denote the rms value of $F(t)$ and $V(t)$.

\section{Wave amplitude power spectrum}
\begin{figure}[t]
\begin{center}
\includegraphics[height=60mm]{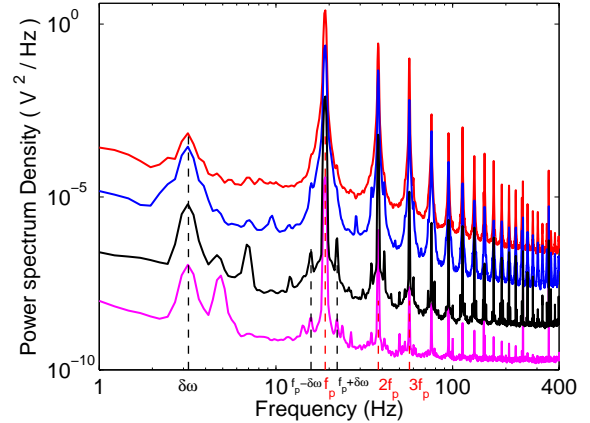}
\caption{(color online). Power spectrum density of wave amplitude for different increasing amplitudes of forcing (from bottom to top). Sinusoidal forcing $f_{p}=19$ Hz. Wave maker immersion $H=10$ mm. Curves are shifted vertically for clarity by a factor 1, 10, 100 and 1000 (from bottom to top). $\delta_\omega=3.2$ Hz (see text).}
\label{mono}
\end{center}
\end{figure}

Figure \ref{mono} shows the power spectrum density of the wave amplitude $S_{\eta}(f)$ for different forcing amplitudes when the wave maker is driven sinusoidally at $f_{p} \simeq f_{gc}$. At low forcing amplitude (bottom curve), the power spectrum exhibits a set of peaks corresponding to the excitation frequency $f_{p}$ and the harmonic ones $nf_{p}$, with $n$ an integer. We observe also the presence of two peaks of small amplitudes at adjacent side-band frequencies $f_p\pm \delta\omega$, and a peak at $\delta\omega$. These latter are due to the Benjamin-Feir instability that destabilizes a monochromatic weakly non-linear wave train in deep-water into a wave train modulated in amplitude with a modulation frequency $\delta\omega$ \cite{BenjaminFeir67}. When the forcing amplitude is increased, the growth of the peak amplitudes at the driving and harmonic frequencies is observed as well as a broadening in frequency. For the maximum forcing amplitude (top curve), the peak at $f_p$ and its harmonics are still observed whereas the peak amplitude at $\delta\omega$ has decreased since the frequency broadening of the fundamental is larger than $\delta\omega$ (curves of Fig.\  \ref{mono}  are shifted vertically for clarity).  Similar results are found when the frequency of the sinusoidal forcing is varied around the crossover frequency $f_{gc}$.  Even for the maximum accessible amplitude of this forcing (top curve in Fig.\ \ref{mono}), the spectrum is still discrete with no power-law spectrum neither in the low frequency part of the spectrum nor in the high frequency part. An easy way to increase the forcing energy at this crossover scale is to use, instead of a sinusoidal forcing, a random one within a narrow frequency bandwidth around $f_{gc}$.

\begin{figure}[t]
\begin{center}
\includegraphics[height=60mm]{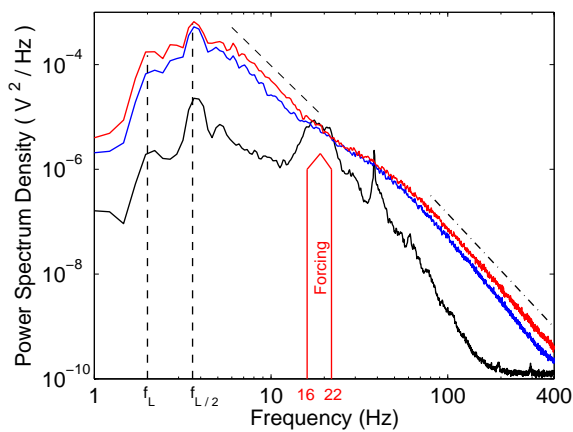}
\caption{(color online). Power spectrum density of wave amplitude for different forcing amplitudes: $\langle I \rangle=2$ mW (lower curve), $\langle I \rangle=15$ mW (medium curve) and $\langle I \rangle=45$ mW (upper curve). Slopes are:  -3.3 (dashed line) and -3.6 (dot-dashed line). Vertical dashed-lines: theoretical resonant frequencies of the vessel $f_{L}=2$ Hz and $f_{L/2}=3.6$ Hz. Maximum amplitude of the spectrum centered at 3.7 Hz is close to the vessel mode $f_{L/2}$. Random forcing: $16 \leq f_p \leq 22$ Hz (thick arrow). Wave maker immersion $H=10$ mm. Vessel size: $L=20$ cm.}
\label{cinv}
\end{center}
\end{figure}

Figure \ref{cinv} shows the power spectrum density of the wave amplitude $S_{\eta}(f)$ for different forcing amplitudes when the wave maker is driven with a band-pass filtered random noise in a frequency range $f_{p} \in [16, 22]$ Hz. At low forcing amplitude (bottom curve), the peak at the forcing frequencies slightly broadens and its harmonic near $2f_{p}$ appears. No power law is observed neither for $f>f_{p}$ nor for $f<f_{p}$. When the forcing amplitude is increased, the spectrum amplitude at $f_{p}$ does not increase whereas the ones at frequencies located on both sides of $f_{p}$ grew significantly. For high enough forcing and for $f<f_{p}$ (corresponding to the gravity wave regime), a power law spectrum is observed from the forcing scale ($f_p$) up to a larger scale over a range less than one decade in frequency (see top curve and dashed line in Fig. \ref{cinv}). Although, quite narrow in frequency, this range corresponds to more than one decade in wave number $k$ using the dispersion relation of linear gravity waves $\omega(k)=\sqrt{gk}$ with $\omega\equiv 2\pi f$, $k\equiv2\pi/\lambda$ and $\lambda$ the wavelength. This power-law spectrum is found to scale as $f^{-\alpha}$ with $\alpha=3.3 \pm 0.2$ and could come from the upscale wave action transfers of gravity wave turbulence.  The gravity wave power spectrum for such an inverse cascade is predicted as $S_{\eta}^{theo}\sim \tilde{q}^{1/3} g f^{-11/3}$ \cite{Zakharov82,ZakharovBook} where $ \tilde{q}$ is the mean wave action flux ($\tilde{q}$ has dimension of [$ \tilde{\varepsilon}$]$/$[$\omega$], i.e. $L^3/T^2$, with $ \tilde{\varepsilon}$ the mean energy flux). The experimental exponent, $\alpha=3.3 \pm 0.2$, found here is thus in rough agreement with the predicted one $11/3 \simeq 3.6$. For $f>f_{p}$ (corresponding to the capillary wave regime), the spectrum has a rounded shape for $f \gtrsim f_{p}$, and displays a power-law for higher frequencies (see top curve and dot-dashed line in Fig. \ref{cinv}). This latter is related to capillary wave turbulence although the power-law exponent found here ($-3.6$) is larger than the expected one ($-17/6$) by weak turbulence theory \cite{ZakharovBook}. The rounded shape of the spectrum linking the inverse cascade of gravity waves and the direct cascade of capillary waves could be due to an accumulation of energy flux in this region. Indeed, for gravity waves, an inverse cascade of wave action is theoretically accompanied with a direct cascade of energy flux \cite{ZakharovBook}.

Note that the power-law of the inverse regime of gravity waves is typically reached 50 s after the starting of the forcing. It thus not allow us to collect enough statistics to accurately studied the transient regime of the dynamical growth of these upscale transfers. 

\section{Discussion}
One can wonder if the above observation of an upscale wave generation could be ascribed to a mechanism different than the inverse cascade one suggested here.

i) Cross waves generated by a wave maker by parametric instability are known to propagate at a frequency half of the driving one \cite{crosswaves}, and could thus polluted the wave spectrum at small scales. However, the onset of occurrence of cross-waves theoretically depends on the amount of water displaced by the wave maker \cite{Garrett70}. Thus, in our experiments, the immersed part of the wave maker has been decreased to reduce the amount of water displaced by this latter. When the immersion depth $H$ of the wave maker is lower than two thirds of the fluid depth $h$ (i.e. $H<12$ mm), we find that the power spectrum of the wave amplitude does not exhibit a peak at half frequency the driving one in contrast with the one observed for a deep plunging wave maker. No cross wave is thus emitted here when $H$ is small enough (as it is the case in all figures).

ii) For our depth of fluid ($h=18$ mm), waves of wavelengths $\lambda \ll 12$ cm are in a deep-water regime ($kh \gg1$). The continuum power law in the gravity regime is observed in Fig.\ \ref{cinv} for $f > 4$ Hz corresponding to wavelengths $\lambda < 8.6$ cm ($kh > 1.3$). One thus probes $kh > 1$. Depth effects on wave turbulence properties have been observed to occur for smaller depths (typically when h < 12 mm for the same frequency range as here) \cite{Falcon11}. The upscale wave generation is thus not related to a depth effect.

iii) The upscale wave generation observed here is not also related to a zero-frequency modulation instability \cite{Punzman} nor to a defect-mediated turbulent regime reported \cite{Falcon09}. In these studies, the amplitude of the wave spectrum at zero-frequency is of the order of the maximum amplitude of the spectrum. Here, the amplitude of the zero-frequency spectrum is negligible compared to the maximum amplitude of the spectrum (see Fig.\ \ref{cinv}). Moreover, our forcing is by means of a wave maker driven with random noise within a frequency band, and not by parametrically vibrated the vessel at a single frequency as in \cite{Punzman,Falcon09}.

Thus, the possible spurious effects listed above are not responsible of the observation of the upscale wave generation that we ascribed to an inverse cascade mechanism. An indirect evidence of this mechanism is the following. If the system is now driven with a forcing at a higher frequency bandwidth in the capillary regime (e.g. $f_{p} \in [30, 35]$ Hz), no power law is observed on the wave spectrum at low frequency as expected by wave turbulence theory \cite{ZakharovBook}. Indeed, capillary wave turbulence is assumed to be generated by a process involving 3-wave interactions whereas inverse cascade is predicted to occur only for an even number of interacting waves (e.g. the 4-wave interaction process of gravity wave turbulence) \cite{ZakharovBook}.

\section{Upscale transfer cut-off}
Let us now focus on the low frequency cut-off of the power law spectrum of the gravity regime. A peak centered at 3.7 Hz is observed in the low-frequency part of the spectrum (maximum amplitude of the spectrum in Fig.\ \ref{cinv}), that seems to end the power law. This frequency corresponds to a wavelength close to half of the vessel size, $L/2$. Indeed, by using the dispersion relation of gravity waves $\omega(k)=\sqrt{gk\tanh{kh}}$, the resonant frequencies of the vessel corresponding to wavelengths $2L$, $L$ and $L/2$ are respectively $1$, 2 and 3.6~Hz. The modes corresponding to wavelengths $L$ and $L/2$ are displayed in Fig.\ \ref{cinv} (see vertical dashed lines), and well correspond to peaks observed on the spectrum. The peak centered at 3.7 Hz is thus related to a resonant frequency mode of the vessel that fixes the lowest accessible frequency of the power-law gravity spectrum. The low frequency cut-off of the inverse cascade is thus due to the horizontal finite size of the experimental setup. To what extent this cut-off is related to a wave condensation phenomenon (large scales generated by the inverse cascade accumulate in a fundamental mode \cite{Korotkevitch08}) remains an open question. Finally, note that the spectrum amplitude at the resonant frequency $f_{L/2}$ increases with the driving but does not broaden in frequency (see Fig.\ref{cinv}). This rules out a possible mechanism that would transfer energy directly from this fundamental mode to smaller scales instead of the suggested mechanism of upscale wave generation by an inverse cascade of wave action. 

\begin{figure}
\begin{center}
\includegraphics[height=60mm]{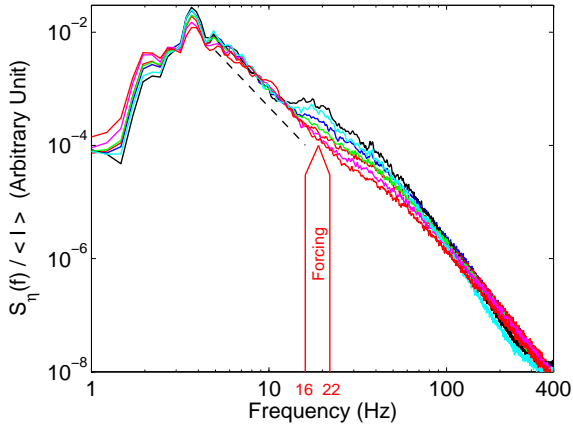}
\caption{(color online). Power spectrum density of wave amplitude rescaled by the mean injected power $S_{\eta}/\langle I \rangle$. Different forcing amplitudes $10 \leq \langle I \rangle \leq 60$ mW and wave maker immersions $H=9, 10$ and 11 mm. Slope of dashed line is -3.3. Random forcing: $16 \leq f_p \leq 22$ Hz.}
\label{spectre}
\end{center}
\end{figure}

\begin{figure}
\begin{center}
\includegraphics[height=60mm]{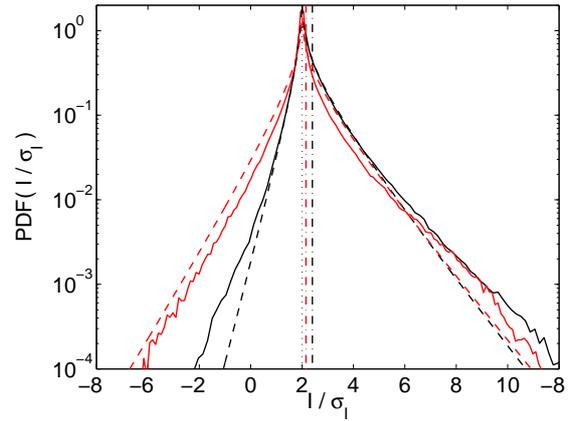}
\caption{(color online). PDFs of the rescaled injected power $I/{\sigma_{I}}$. Red (light gray) solid line: with an {\it inverse} gravity cascade (intermediate scale forcing $f_{p} \in [16,22]$ Hz). Black solid line: with a {\it direct} gravity cascade (large scale forcing $f_{p} \in [0.1,6]$ Hz). $\langle I \rangle=45$ mW for both cases. Dashed lines are the predictions of Ref. \cite{Falcon08} with $\langle I \rangle/\sigma_{I}=0.15$ [red (light gray)], and 0.5 (black). Dashed-dotted lines are $\langle I \rangle/\sigma_{I}=0.15$ [red (light gray)], and 0.5 (black). $H=10$ mm.}
\label{pdf}
\end{center}
\end{figure}

\section{Spectrum scaling with the injected power}
We now turn to the scaling of the spectrum amplitude $S_{\eta}(f)$ with the mean injected power $\langle I \rangle$.  Figure \ref{spectre} shows the wave spectra $S_{\eta}(f)$ rescaled by $\langle I \rangle$ for different $\langle I \rangle$ (varying by a factor 6) and for a random forcing at intermediate scales ($f_{p}\in [16,22]$ Hz). For frequencies $f<f_{p}$, all spectra are found to collapse. The best scaling for this regime is $S_{\eta} \sim \langle I \rangle^{1 \pm 0.1}$. Surprisingly, this scaling for the {\it inverse} regime of gravity wave turbulence is the same as the one obtained for both gravity and capillary {\it direct} cascades when using a large scale forcing \cite{Falcon07a,Punzman}. We have checked that the immersion height of the wave maker $H$ does not change this scaling of both {\it direct} cascades for the large scale forcing. Finite size effects of the container could be an explanation for this similar scaling for both inverse and direct gravity wave regimes: inverse transfers generate larger and larger scales up to almost the vessel size leading to a significant finite size effect, as when the forcing is at large scale since it directly generates scales of the order of the vessel size. The inverse cascade of gravity waves theoretically depends on the mean wave action flux $ \tilde{q}\equiv \int q(k)dk$ and its spectrum is predicted to scale as $ \tilde{q}^{1/3}$ (see above).  Although  the mean energy flux $ \tilde{\varepsilon}\equiv \int \varepsilon(k)dk$ is a measurable quantity ($ \tilde{\varepsilon} \sim \langle I \rangle$ - see \cite{Falcon07a}), and $q(k)$ and $\varepsilon(k)$ are related by $\frac{\partial \varepsilon(k)}{\partial k} = \omega(k)\frac{ \partial q(k)}{\partial k}$, we have currently no way to measure $ \tilde{q}$. An estimate of  the value of $\tilde{q}$ would require an estimate of the number of wave modes in interaction. It is not possible with a temporal measurement at a single location but should deserves further studies with a spatio-temporal one \cite{Herbert2010}.

\section{Distribution of the injected power}
Figure\ \ref{pdf} shows the probability density function (PDF) of the instantaneous injected power, $I(t)$, rescaled by its rms value, $\sigma_{I}$, in the presence of an inverse cascade (intermediate scale forcing) or in the presence of a direct gravity cascade (large scale forcing) both for the same forcing value of $\langle I \rangle$. In both cases, the main properties of the PDF found in \cite{Falcon08} are observed: i) the most probable value is 0, ii) strong fluctuations occur up to 10 times the rms value $\sigma_{I}$, iii) the PDF displays roughly exponential tails for both positive and negative values of $I$, and iv) the model introduced in Ref.~\cite{Falcon08} with any adjustable parameter (when $\langle I \rangle/\sigma_{V}\sigma_{F}\simeq \langle I \rangle/\sigma_{I}$ is experimentally known) captures well both PDF shapes (see dashed lines in Fig.\ \ref{pdf}). Moreover, we find that both PDFs do not depend on the forcing amplitude. Indeed, the PDF($I/\sigma_{I}$) for different forcing (within our range of $\langle I \rangle$) well collapse either for the direct cascade case or for the inverse one (not shown).

The main difference between both PDFs is that negative events of injected power are more likely when upscale transfer occurs than with only direct gravity cascade. It means that the energy coming back to the wave maker is more probable than for a direct cascade leading to a more symmetric PDF. This increase of energy transferred to the wave maker can be seen as a large scale effective dissipation.

\section{Conclusion}
Experiments related to the observation of an inverse cascade in wave turbulence are scarce. Here, we have performed experiments in order to observe in laboratory the inverse cascade in gravity wave turbulence on the surface of a fluid. To wit, surface waves are forced at an intermediate scale corresponding to the gravity-capillary wavelength. Due to nonlinear wave interactions, waves at larger scales are observed and their amplitudes exhibit a frequency-power law spectrum at high enough forcing. Both observations are ascribed to the upscale wave action transfers of gravity wave turbulence. The exponent of the spectrum is close to the predicted value of the weak turbulence theory. The amplitude of the spectrum is found to scale linearly with the mean energy flux. Finally, the measurement of the probability distribution of injected power fluctuations shows that the energy given back to the wave maker is more likely in the presence of upscale (inverse) transfers than in the presence of direct cascade in gravity wave turbulence.

\begin{acknowledgments}
We thank M. Berhanu for fruitful discussions. This work has been supported by ANR Turbonde BLAN07-3-197846.
\end{acknowledgments}

\end{document}